\documentclass[sn-basic]{sn-jnl}% Math and Physical Sciences Reference Style

\jyear{2025}%

\usepackage{amsmath} %usually use this to have vectors in italics etc.
\usepackage{sidecap}
\usepackage{lineno}
\usepackage{relsize}
\theoremstyle{thmstyleone}%

\theoremstyle{thmstyletwo}%

\theoremstyle{thmstylethree}%
\def\narrower{%
  \advance\leftskip\parindent
  \advance\rightskip\parindent}

\begin{document}

    \title[Amalgamations]{\centering Amalgamations in a hierarchy \\ \centering as a way of variable selection \\ \centering in compositional data analysis \\ \vspace{0.3cm}}

\author*[1]{\fnm{Michael} \sur{Greenacre}}\email{michael.greenacre@upf.edu}
\author[2]{\fnm{Martin} \sur{Graeve} }\email{martin.graeve@awi.de}

\affil*[1]{\orgdiv{Department of Economics and Business, and Barcelona School of Management}, \orgname{Universitat Pompeu Fabra}, \orgaddress{\street{Ramon Trias Fargas, 25--27}, \city{Barcelona}, \postcode{08005}, \country{Spain}}}

\affil[2]{\orgdiv{Ecological Chemistry}, \orgname{Alfred Wegener Institute for Polar and Marine Research}, \orgaddress{\street{Am Handelshafen, 12}, \city{Bremerhaven}, \postcode{27570}, \country{Germany}}}

%\author*[1]{\fnm{} \sur{}}\email{}
%\author[2]{\fnm{} \sur{} }\email{}

%\affil*[1]{\orgdiv{, \orgaddress{\street{}, \city{}, \postcode{}, \country{}}}}

%\affil[2]{\orgdiv{}, \orgname{}, \orgaddress{\street{}, \city{}, \postcode{}, \country{}}}

%%==========%%
%% Abstract %%
%%==========%%

\abstract{In certain fields where compositional data are studied, the compositional components, called parts, can be combined into certain subsets, called amalgamations, which are based on domain knowledge. Furthermore, these subsets can form a natural hierarchy of amalgamations subdividing into sub-amalgamations.
The authors, a statistician and a biochemist, demonstrate how to create a hierarchy of amalgamations in the context of fatty acid compositions in a sample of marine organisms. Following a tradition in compositional data analysis, these amalgamations are transformed to logratios, and their usefulness as new variables is quantified by the percentage of total logratio variance that they explain. This method is proposed as an alternative method of variable selection in compositional data analysis.}


\maketitle

Manuscript version: \today

%%%%%%%%%%%%%%%%%%%%%%
\section{Introduction} \label{Introduction}
Compositional data are nonnegative multivariate data that are generally expressed as proportions of a whole, called \textit{compositions}. 
Examples are sets of counts or other features such as monetary values, physical quantities on the same scale, hours spent on certain tasks, and -- in the present paper -- fatty acid components in biochemistry.
In each of these cases, the multivariate observations are usually expressed relative to their respective totals, since they carry relative rather than absolute meaning.
In general, the features/components/constituents of a composition are called its \emph{parts}.
John Aitchison's pioneering work in compositional data analysis, called CoDA \citep{Aitchison:82, Aitchison:86}, was aimed primarily at incomplete compositions, where the parts included in the study were not all that could have been observed.
The fact that these incomplete compositions had been normalized to sum to 1 created spurious correlations between the parts, and so the use of \emph{subcompositionally coherent} ratios between parts was advocated.
Two different biochemistry laboratories arriving at different fatty acid compositions, for example, would have a subset of parts (fatty acids) in common, but their relative values in the composition would not be comparable.
Ratios between pairs of parts of this subset, however, could be compared across studies, since they are invariant to which parts have been included.  

Because of the skewness and multiplicative nature of ratios, they are preferably log-transformed, leading to \emph{logratios}, for example the logratios log(A/B) between all pairs (A,B) of parts.
These pairwise logratios (PLRs) form the backbone of any analysis of incomplete compositional data.

Groups of parts are naturally combined through \emph{amalgamation}, that is, summing their values.
For example, the group of saturated fatty acids (SFA) is combined by summing all the fatty acid values with labels X:0, e.g. SFA = 14:0 + 15:0 + 16:0 + ... etc.
While this addition of parts might seem self-evident, it is surprising that some researchers actually combine parts using geometric means to satisfy a mathematical nicety that has no bearing on common practice or common sense.
For example, in their book on compositional data analysis \cite{PawlowskyEtAl:15} state explicitly that ``Amalgamation is incompatible with the techniques presented in this book", and dismiss the use of ratios of amalgamations, basing this view on a purely mathematical argument.  
By contrast, \cite{Aitchison:82, Aitchison:86} promoted the use of amalgamated parts, as also did \cite{ScealyWelsh:14}.
In their compositional data analysis tutorial, \cite{Smithson:22} say  ``The geometric mean of compositional parts cannot be interpreted as simply combining the parts, but their sum can."
In this article we adhere to the latter view, which, as we say, is self-evident in every field where compositional data are found, and especially in the field of biochemistry.
For further commentary on the combining of parts in CoDA, see \cite{Greenacre:19d}

Amalgamations of fatty acids (FAs) in biochemistry are already implemented in the three major classifications of saturated FAs (SFA), monounsaturated FAs (MUFA) and polyunsaturated FAs (PUFA).
Beneath these three major types, further subdivisions can be naturally defined, forming a divisive dendrogram. %until the individual fatty acids are reached at the lower end of the hierarchy.
The objective of this paper is to quantify how much this hierarchical structure explains the \emph{total logratio variance}, which is a measure of how much information there is in a FA compositional data set. 
This total quantity is based on the variances of all the PLRs in the data set.
A hierarchy of amalgamations can be created purely on statistical criteria, using a clustering algorithm  \citep{Greenacre:20}.
In the present work, however, amalgamations are preferably defined based on domain expertise, by a biochemist in this case, creating the hierarchical structure step by step.
At each subdivision of the parts into groups (``sub-amalgamations"), the percentage of logratio variance explained by the logratios of the amalgamated parts will be calculated. 
At a certain stage there is no natural grouping of the parts that is suggested by domain knowledge, and then the PLRs of those parts could be used to explore the unexplained logratio variance.
The main objective of this exercise is a type of dimension reduction in the form of variable selection, to arrive at a small set of amalgamated parts, with clear substantive meaning, whose logratios explain a sufficiently large percentage of the total logratio variance, and which can thus replace the many individual FAs for downstream analysis. 

%%%%%%%%%%%%%%%%%%%%%%%%%%%%%%%%%%%%%%%%%%%%%%%%%%%%%%%%

\section{Material and methods}
\subsection{Data set \texttt{\textbf{FattyAcids}}}
Calanoid copepods are small marine crustaceans, and the present samples were collected during an extensive field study in Rijpfjorden, a high Arctic sea ice dominated ecosystem, during the International Polar Year 2007/2008. The seasonal development (winter--spring--summer transition) of the key pelagic grazer was investigated together with the ice algae and phytoplankton growth, see \cite{Soreide:10}. This dataset is composed of 42 samples of the Arctic \emph{Calanus} species \emph{Calanus glacialis}. The samples were analysed for fatty acid\footnote{Fatty acids, affectionately called the ``Fats of Life", are denoted by their chemical structure. For example, 20:5(n-3) has a chain of 20 carbon atoms with 5 double carbon bonds, and its last double bond is 3 carbons away from the methyl, or omega, end (i.e., it is an omega-3 FA). Saturated FAs have no double bonds, e.g., 14:0; monounsaturated FAs have one double bond, e.g., 16:1(n-7); and polyunsaturated FAs have two or more double bonds, e.g., 16:2(n-4), 18:4(n-3). Short chain MUFAs have up to 18 carbon atoms, while long chain MUFAs have 20 or more carbon atoms. For correct chemical nomenclature, see IUPAC Nomenclature of Organic Chemistry, Sections A, B, C, D, E, F and H, 1979 Edition, Pergamon Press, 1979.} compositions to reveal lipid and fatty acid physiology and to conclude their nutritional status. 
Gas chromatography provided a set of 40 fatty acids (FAs) on each sample, which are included in the statistical analysis.
For an overview of the use of FAs in marine biochemistry, see \cite{Dalsgaard:03}.
A small part of the data set is given in Table \ref{FattyAcid_Ddata} and the full data set is available online -- see the Data Availability section at the end of the paper. 

\begin{table}[h]
\vspace{-0.1cm}
\caption{Part of the data set \texttt{\textbf{FattyAcids}}, showing percentages for the first six rows (out of 42), and the first six columns (out of 40). The data are shown correct to three decimals and the row sums are all equal to 100.
\label{FattyAcid_Ddata}}
\hspace{0.8cm}
\begin{tabular}{lrrrrrrc}
 {\sf Season} & {\sf\em 14:0} & {\sf\em 14:1(n-5)} & {\sf\em i-15:0} & {\sf\em a-15:0} & {\sf\em 15:0} & {\sf\em 15:1(n-6)} & $\cdots$ \\
\hline\\[-6pt]
{\sf winter} & 13.854 &  0.203 & 1.190 & 0.446 & 0.847 & 0.103 & $\cdots$ \\
{\sf winter} & 11.827 &  0.148 & 1.236 & 0.460 & 1.056 & 0.084 & $\cdots$ \\ 
{\sf winter} &  6.457 &  0.000 & 0.768 & 0.248 & 0.514 & 0.000 & $\cdots$ \\
{\sf winter} & 12.011 &  0.148 & 1.130 & 0.411 & 0.796 & 0.121 & $\cdots$ \\ 
{\sf summer} &  6.579 &  0.193 & 0.329 & 0.000 & 0.264 & 0.000 & $\cdots$ \\
{\sf summer} &  6.704 &  0.193 & 0.315 & 0.000 & 0.254 & 0.000 & $\cdots$ \\[-2pt]
\ \ \ \ \ $\vdots$ & $\vdots$  \ \ \ \ & $\vdots$  \ \ \ \ & $\vdots$  \ \ \ \ & $\vdots$  \ \ \ \ & $\vdots$  \ \ \ \ & $\vdots$  \ \ \ \ & \ \ $\vdots$ \ \  \\
\hline
\end{tabular}
\vspace{-0.2cm}
\end{table}

\subsection{Methods}
The method is now described for the creation of a hierarchical structure of amalgamated parts, FAs in this case. 
Statistical concepts and procedures are used to quantify how well the amalgamated parts, specifically their logratios, explain the total logratio variance, which is the measure of information in the whole data set.
We first define total logratio variance, then the method of redundancy analysis that will allow us to measure explained variance, and finally the way the amalgamations are created.

The zeros in the data set need to be replaced by small positive numbers to make possible the the calculation of logratios. 
We chose one of the simplest methods, replacing the zeros by two-thirds of the smallest positive number in the respective column \citep{Greenacre:21, Lubbe21}. 

\medskip
\leftline{\textit{\textbf{Total logratio variance}}}
\smallskip

\noindent
For a $J$-part composition, there are $J(J-1)/2$ pairs of parts and thus $J(J-1)/2$ unique pairwise logratios PLRs, defined as follows.
Given a composition $\{\, x_1, x_2, \ldots, x_J\,\}$ of positive values, the unique set of PLRs is defined as
\vspace{-0.2cm}
\begin{eqnarray}
{\rm PLR}(j,k) &=& \log(x_j/x_k) ,\quad j, k =1,2,\ldots,J, \,\, j < k.    \label{PLR}\\ 
              &=& \log(x_j) - \log(x_k) . \nonumber
\end{eqnarray}

\vspace{-0.2cm}
\noindent
(notice that it does not matter technically which parts are the numerator and the denominator, and these can be interchanged for purposes of interpretation).
In the present application $J=40$, so there are $40\times 39 /2 = 780$ unique pairwise logratios.
Each of these PLRs has a variance, denoted by ${\sf var}_{jk}$ with a double subindex, which is computed by the usual formula except that the sum of squared deviations from the mean is divided by the number of samples $I$ (in this application, $I=42$) and not by $I-1$.

%Notice that the summation involves ${\small\frac{1}{2}} J (J-1)$ logratio variances, but the variances are not averaged but rather divided by $J^2$.
%This is because $1/J$ is regarded as the weight assigned to each of the $J$ parts, and each logratio variance is multiplied by the product of the weights of its two parts which is $1/J^2$.
We give the most general definition of the total logratio variance, abbreviated as \textsf{TotLogVar} where the $J$ parts have pre-assigned positive weights $c_j, j=1,\ldots,J$, where $\sum_j c_j=1$ (thus, the weights are themselves a composition). 
The definition of the {\it total weighted logratio variance} is:
\vspace{-0.1cm}
\begin{equation}
\label{totalvariance}
  {\sf TotLogVar} = \sum_{j<k} c_j\, c_k\,{\sf var}_{jk}
\end{equation}

\vspace{-0.2cm}

\noindent
If the parts are not differentially weighted, then $c_j = \frac{1}{J}$ for all $j$ and (\ref{totalvariance}) reduces to ${\sf TotLogVar} = \frac{1}{J^2}\sum_{j<k} {\sf var}_{jk}$.

Computing the total variance based on all logratios requires the computation of $\frac{1}{2} J (J-1)$ variances.
However, exactly the same result can be obtained by computing the $J$ variances of the centred logratios (CLRs), which is a substantial computational advantage. 
There are $J$ CLRs defined as follows, in the weighted form: 
\vspace{-0.2cm}
\begin{equation}
  \label{CLR}
  \textrm{CLR}(j) \ = \ \log\left(\frac{x_j}{\prod_k x_k^{c_k}}\right) \ = \ 
                  \log(x_j)-\sum_{k=1}^J c_k\,\log(x_k), \quad j=1,\ldots,J
\end{equation}
\vspace{-0.2cm}

\noindent
that is, CLRs are the log-transformed parts centred with respect to their (weighted) mean across the parts.
For unweighted CLRs, centre by the arithmetic mean $(1/J)\sum_k \log(x_k)$.
The variances of the CLRs, that is, the columns of the row-centred log-transformed compositional data matrix, have variances denoted by ${\sf var}_j,\, j=1,\ldots,J$, with a single subscript, where variances are computed once more by dividing the sum-of-squared deviations from the mean of each CLR by $I$, not $I-1$.
%Then the total weighted variance is equal to $(1/J)\sum_j {\rm var}_j$.
%For the more general weighted version, the CLRs have first to be recomputed with respect to weighted row means of the log-transformed table before computing their variances.
The weighted version of {\sf TotLogVar} in (\ref{totalvariance}) is then equivalently computed as 
\vspace{-0.2cm}
\begin{equation}
\label{totalvarianceCLR}
  {\sf TotLogVar} = \sum_j c_j\, {\sf var}_{j}
\end{equation}
Again, the unweighted case is the special case of (\ref{totalvarianceCLR}) when $c_j = 1/J$ for all $j$: ${\sf TotLogVar} = \frac{1}{J} \sum_j {\sf var}_{j}$, where the CLRs are centred by regular arithmetic averages before computing the variances. 
Using the CLRs saves computational effort, since only $J$ variances are computed, not ${\frac{1}{2}} J (J-1)$ variances.
For more details about logratios, logratio variance and weighting, see \cite{Greenacre:18}.

\medskip
\leftline{\textit{\textbf{Redundancy analysis}}}
\smallskip

\noindent
Redundancy analysis (RDA) \citep{Wollenberg:77} is a multivariate version of regression, combined with dimension reduction, where the responses are a vector 
of variables. 
Our use of it here is mostly in multivariate regression, where the responses are the $J$ variables of the CLR-transformed compositional data.
If the $J$ CLRs are gathered in a vector $\bf Y$ ($J\times 1$), then RDA is effectively fitting the linear model:

\vspace{-0.2cm}
\begin{equation}
    {\bf Y} = {\mathbf \alpha} + {\mathbf \beta}_1 \circ {\bf X}_1 + {\mathbf \beta}_2 \circ {\bf X}_2 + \cdots + {\bf E} 
    \label{RDA}
\end{equation}

\smallskip

\noindent
where $\mathbf \alpha$, ${\mathbf \beta}_1$, ${\mathbf \beta}_2$, $\ldots$, are $J\times 1$ coefficient vectors, ${\bf X}_1$, ${\bf X}_2$, $\ldots$, are $J\times 1$ vector variables that are explaining $\bf Y$, $\bf E$ is a vector of residuals, and $\circ$ is the Hadamard product that performs elementwise multiplication.  
In the present application the explanatory variables will be the logratios of the constructed amalgamations, and the RDA will quantify the explained variance of the $J$ variables in $\bf Y$.
The same result would be obtained if $J$ separate multiple regressions were performed, regressing each response in $\bf Y$ on the explanatory variables, and accumulating the explained variances from each regression.
RDA saves doing each one of them and gives the overall explained variance of all the response variables in a single summary.

\medskip
\leftline{\textit{\textbf{Creation of amalgamations}}}
\smallskip

\noindent
The choice of the amalgamations is entirely based on domain knowledge, with the statistical methods supporting the choice.
In addition, later in the process, once there seems to be no more substantively interesting amalgamations to create, the parts in each of the final amalgamations may be analysed by selecting PLRs to further increase explained variance. 
This would be a final exploratory exercise, subject to biochemical interpretation of the results.

The process starts with the creation of major groups of parts, in this case the three amalgamations SFA, MUFA and PUFA, which are now treated as a three-part composition.
To quantify how much of the total logratio variance is explained by this amalgamation composition, the stepwise PLR selection of these three amalgamated parts is performed, as described by \cite{Greenacre:19}. 
These pairwise logratios of amalgamated parts are called \emph{amalgamated} or \emph{summated logratios} (SLRs) \citep{GreenacreGrunskyBaconShone:20, Greenacre:20}.
Because of the linear relationship between the three SLRs, only two SLRs are required to be selected, one less than the number of amalgamations. 
The two SLRs of these parts will be explanatory variables ${\bf X}_1$ and ${\bf X}_2$ in (\ref{RDA}).
Then, within each of these three amalgamations, further subdivisions are proposed by the domain expert, forming SLRs of these ``sub-amalgamations", and quantifying their explained variances using RDA.
In each case, the explanatory variables are accumulated. 
In other words, the RDA has the first two SLRs from the  first set of amalgamations as well as the SLRs of the sub-amalgamations. 
The choice of which SLRs of the sub-amalgamations to choose will again depend on which adds the most explained variance.
Then the SLRs of the other sub-amalgamations are introduced in the same way, by adding their SLRs to those already selected and choosing the one that maximizes explained variance.

%The added explained variance at each step can be visualized in a type of \emph{scree plot} that facilitates the selection of where to cut the hierarchy and arrive at a small set of amalgamated parts, and possibly some single unamalgamated ones as well, to use for further statistical analysis involving much fewer parts than the original set, but capturing most of the variance.

At the end of the process when no more amalgamations seem feasible, it is the parts in each of the lower branches of the hierarchy that can define PLRs amongst themselves.
As an exploratory final step, these PLRs can be explored using a similar stepwise approach proposed by \cite{Greenacre:19}.

%%%%%%%%%%%%%%%%%%%%%%%%%%%%%%%%%%%%%%%%%%%%%%%%%%%%%%%%%%%%%%%%
\section{Results}
As already explained, the first and most evident FA amalgamations to choose are the groupings SFA, MUFA and PUFA, which are logical groups according to the number of double bonds. 
But it also splits according to the physiological nature of the FAs. 
SFA and PUFA are found more in the organism's membranes and MUFA is more connected to the storage of lipids (i.e., fats).

\begin{center}
\begin{figure}[t]
\vspace{-0.3cm}
\centering
\includegraphics[width=12cm]{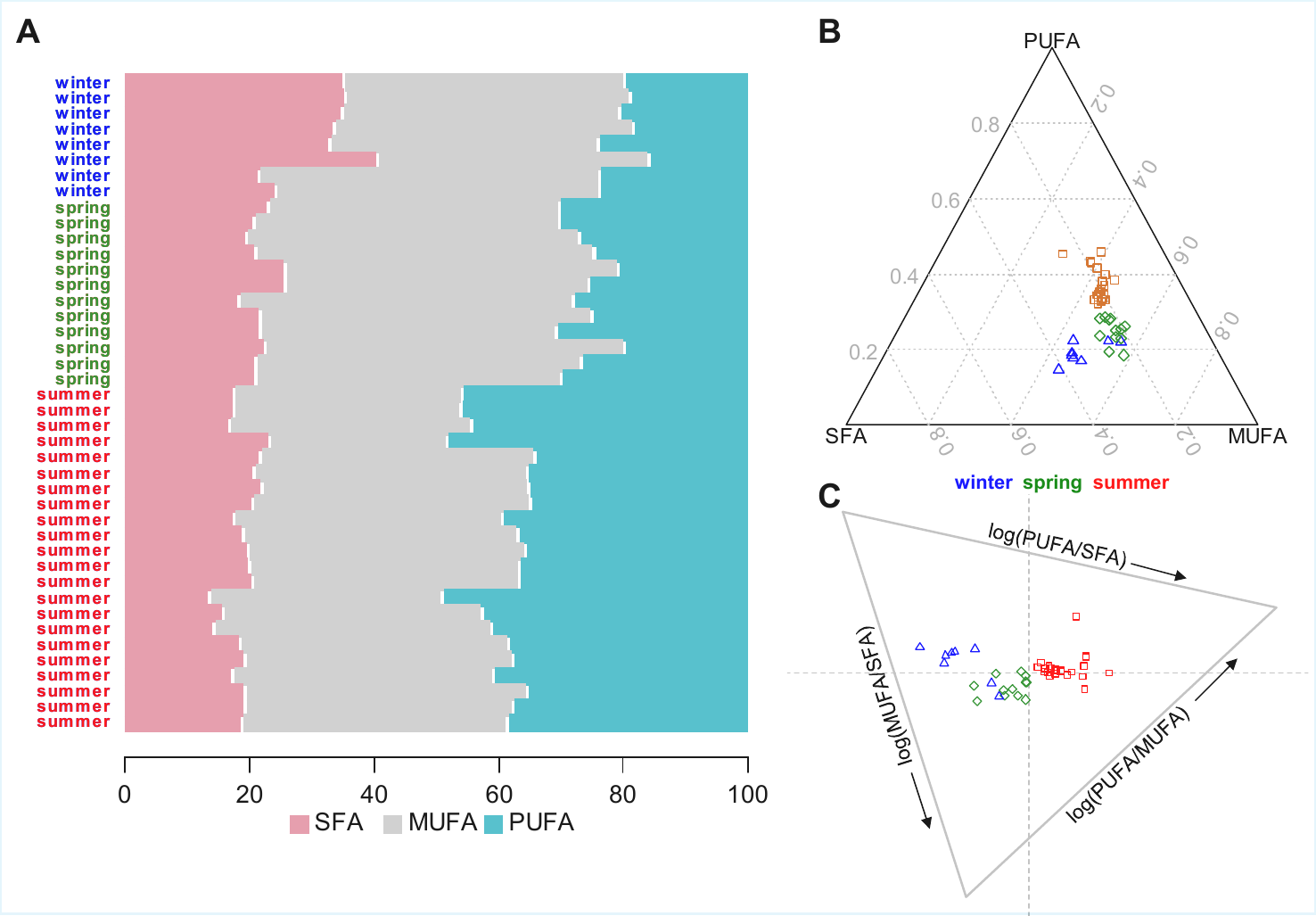}
\caption{A. Compositional barplots of SFA, MUFA and PUFA amalgamations, with labelling of the samples by season. B. Ternary plot in triangular coordinates. C. Logratio analysis (LRA, i.e., PCA of the PLRs) oriented to principal axes. In this exact biplot, 86.3\% logratio variance of the three amalgamations is explained on the horizontal first dimension, and 13.7\% on the second.}
\label{BarTernaryLRA}
\end{figure}
\end{center}

Thanks to the fact that these are three amalgamations, they can be visualized in several ways, as shown in Fig.~\ref{BarTernaryLRA}.
A simple compositional bar plot is shown in Fig.~\ref{BarTernaryLRA}A, where it can already be seen that the component of PUFA is increasing as winter moves to summer, with a compensating decrease of SFA. 
We can already guess that the ratio PUFA/SFA will be important. 
Indeed, the ternary plot in Fig.~\ref{BarTernaryLRA}B shows the winter--spring--summer transition moving from the SFA corner to PUFA. 
Fig.~\ref{BarTernaryLRA}C is a principal component analysis (PCA) (for a recent review, see \cite{GreenacreEtAlPCA:22}), where the three sides depicting logratios (SLRs in this case) are still in a perfect equilateral triangular shape, thanks to the exactness of the biplot visualization.
The horizontal first dimension of this PCA explains 86.3\% of the logratio variance of the three amalgamations, with which log(PUFA/SFA) is the most aligned.  
The PCA of CLRs or SLRs is called logratio analysis (LRA) \citep{Greenacre:18}, a dimension-reduction method which in this three-part case is exactly two-dimensional.

\vspace{-0.4cm}

\begin{center}
\begin{figure}[t]
\vspace{-0.1cm}
\centering
\includegraphics[width=11.5cm]{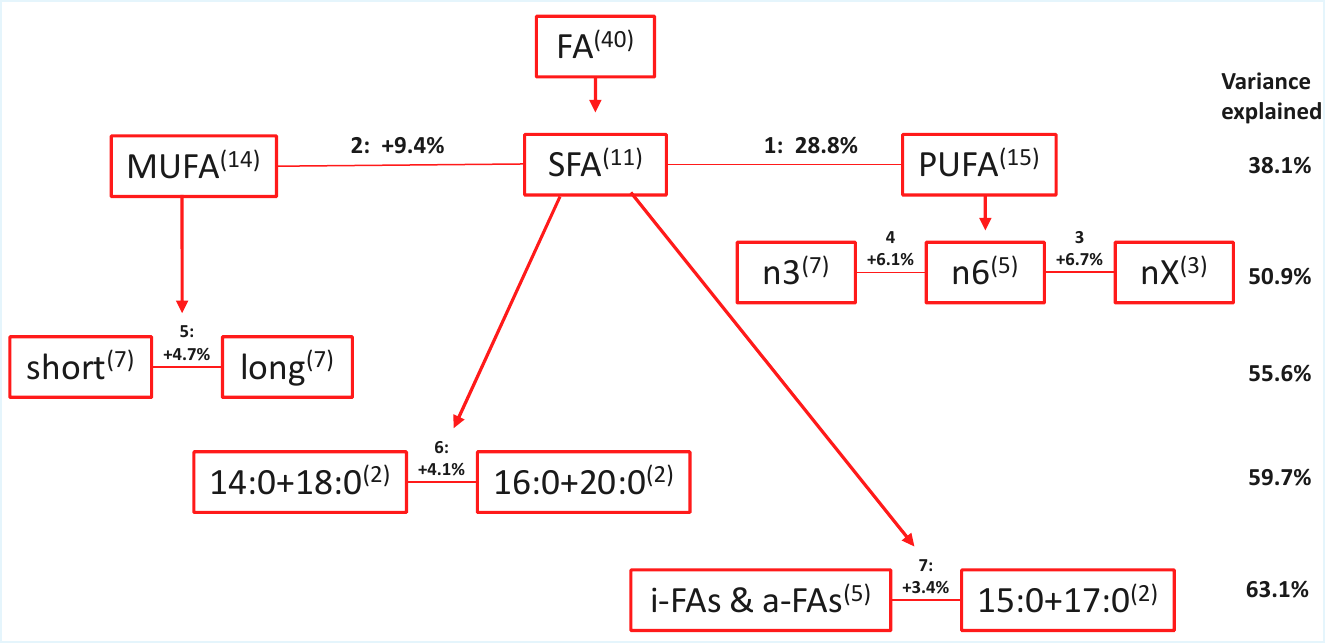}
\caption{Hierarchical summary of amalgamation creation and computation of explained logratio variance by amalgamation (or summated) logratios (SLRs). The amalgamation label is superscripted with the number of included FAs, and the step sequence number is indicated on the connections that define the SLRs linking a pair of amalgamations  A total of 12 amalgamations are defined, and 7 SLRs. The 7 SLRs explain 63.1\% of the total logratio variance based on the 40 FAs. }
\label{FinalResults}
\end{figure}
\end{center}

Our object, however, is to identify specific transformations that maximize explained variance, so we look for the best explanatory SLR, and it turns out, as suspected, to be log(PUFA/SFA), explaining 28.8\% of the variance.
Choosing either log(PUFA/MUFA) or log(MUFA/SFA) as the second SLR gives the same increase in variance explained. 
We choose the latter SLR, which brings the variance explained to 38.1\%, an additional 9.3\%.
These results are summarized in the first row of the summary of the results in Fig.~\ref{FinalResults}.

The next most natural subdivision from a biochemical point of view is to subdivide the PUFAs.
PUFAs are essential for physiological processes such as membrane development. Copepods ingest PUFAs when feeding on a marine algal diet. 
The PUFAs are split according to the position of the double bond, into three sub-amalgamations, the first two of which we denote by n3 (18:3(n-3),18:4(n-3), 20:3(n-3), etc., 7 FAs) and n6 (18:2(n-6), 18:3(n-6), 20:2(n-6), etc., 5 FAs). 
This leaves three PUFAs that are algal markers (16:2(n-4), 16:3(n-4), 16:4(n-1)), which we amalgamate into a group labelled nX (Fig.~\ref{FinalResults}).  
Since the SLR log(n3/n6) is the most important biochemically, it is first added to the explanatory variable set, explaining an additional 6.0\% of the variance, and then the SLR log(n3/nX), explaining an additional 6.8\%, bringing the variance explained up to this point, with only four SLRs defined, to 50.9\%.

The next step is to subdivide the MUFA group, into short and long chain (see footnote in Section 2.1), seven FAs in each case. 
This divides the recently synthesized short-chain FAs from those FAs generally found in storage lipids.
Here there is only one SLR, and this adds an additional 4.7\% explained variance, bringing the explanation up to 55.6\%.

Then it is the turn of the SFA group to be subdivided. There are two pairs of amalgamations that suggest themselves.
The first is 14:0 and 18:0 versus 16:0 and 20:0.
The first subdivison is chosen, because of newly synthesized (\emph{de novo}) FAs. 
The second group might also have a proportion of material that enters the FA pool by degradation of storage FAs.
The resulting SLR adds 4.1\% explained variance.
The second is the i-FAs and a-FAs versus 15:0 and 17:0.
The differentiation of the branched i-FAs and a-FAs (iso and anteiso FAs) versus the group of odd-chain FAs (15:0 and 17:0) was chosen because of the potential origin of the material. 
Branched FAs are known as degradation end products of algal origin, while odd-chain FAs are connected to bacterial input.
The resulting SLR adds 3.4\% explained variance to that already explained by the short- versus long-chain SLR.
Together, these two SLRs in the SFA group account for an additional 7.5\% explained variance, bringing the explained variance accumulated up to this point to 63.1\%.

\vspace{-0.4cm}

The selections made up to now are visually summarized in Fig.~\ref{FinalResults}. 
There are no further evident sub-amalgamations that are suggested, based on biochemical expertise, so the amalgamation process stops here. 
The 7 SLRs, involving 12 amalgamations, have explained 63.1\% of the total logratio variance of 40 FAs.

\begin{center}
\begin{figure}[t]
\vspace{-0.5cm}
\centering
\includegraphics[width=5.4cm]{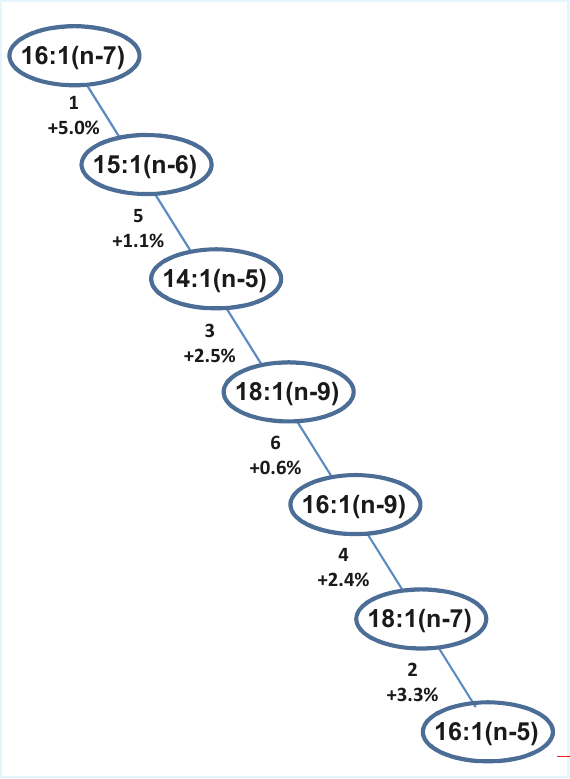}
\caption{Acyclic connected graph of the six PLRs that are formed from the seven short-chain MUFAs on the left of Fig.~\ref{FinalResults}.}
\label{GraphSHortMUFA}
\end{figure}
\end{center}

The amalgamation selection, guided by the biochemist,  stops here.
But there could be further simple PLRs, chosen statistically, that might add several percentage points more to the explained variance and have a substantively interesting biochemical interpretation.
For example, the seven PLRs formed from the short-chain FAs in the MUFA group, on the left of Fig.~\ref{FinalResults}, can be subjected to logratio selection as proposed by \cite{Greenacre:19} and the results are shown visually in Fig.~\ref{GraphSHortMUFA}.
The seven PLRs necessarily form an acyclic connected graph \citep{Greenacre:18}, acyclic because there is no cycle in the graph, and connected because all seven FAs are linked.
The first PLR that enters is log(16:1(n-7)/15:1(n-6)) (shown at the top of the graph, labelled 1, and adds 5.0\% to the explained logratio variance, more than most of the SLRs in Fig,~\ref{FinalResults}.
The second best PLR is log(18:1(n-7)/16:1(n-5)) (bottom of the graph, labelled 2), adding 3.3\% explained variance, and so on.
These top two logratios involve trace FAs, and there seems to be no substantive biochemical interpretation for them.
Hence, we can conclude that the branch of short-chain MUFAs is not worth subdividing further.
The same exercise can be conducted on the other amalgamations at the foot of the hierarchy.

\vspace{-0.7cm}
\begin{center}
\begin{figure}[H]
\centering
\includegraphics[width=9cm]{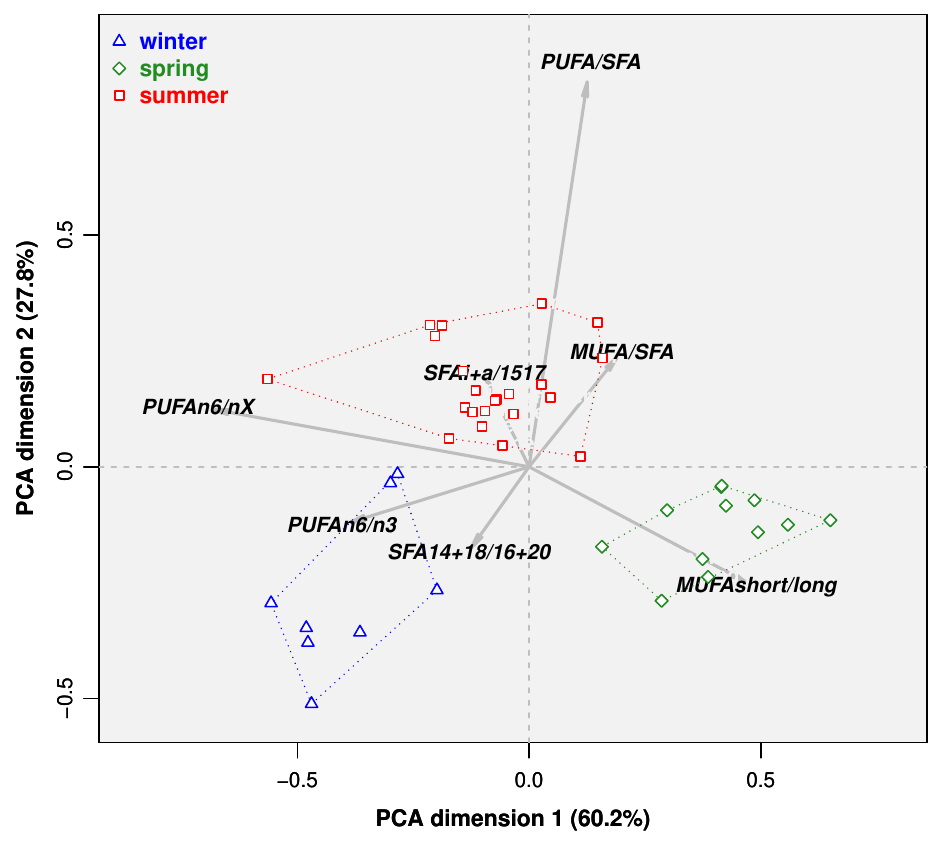}
\caption{PCA of the 7 SLRs created by the hierarchical variable selection process, showing the convex hulls of the three seasons' points and their perfect separation.}
\label{FinalPCA}
\end{figure}
\end{center}

\vspace{-1cm}

\begin{center}
\begin{figure}[H]
\centering
\includegraphics[width=9cm]{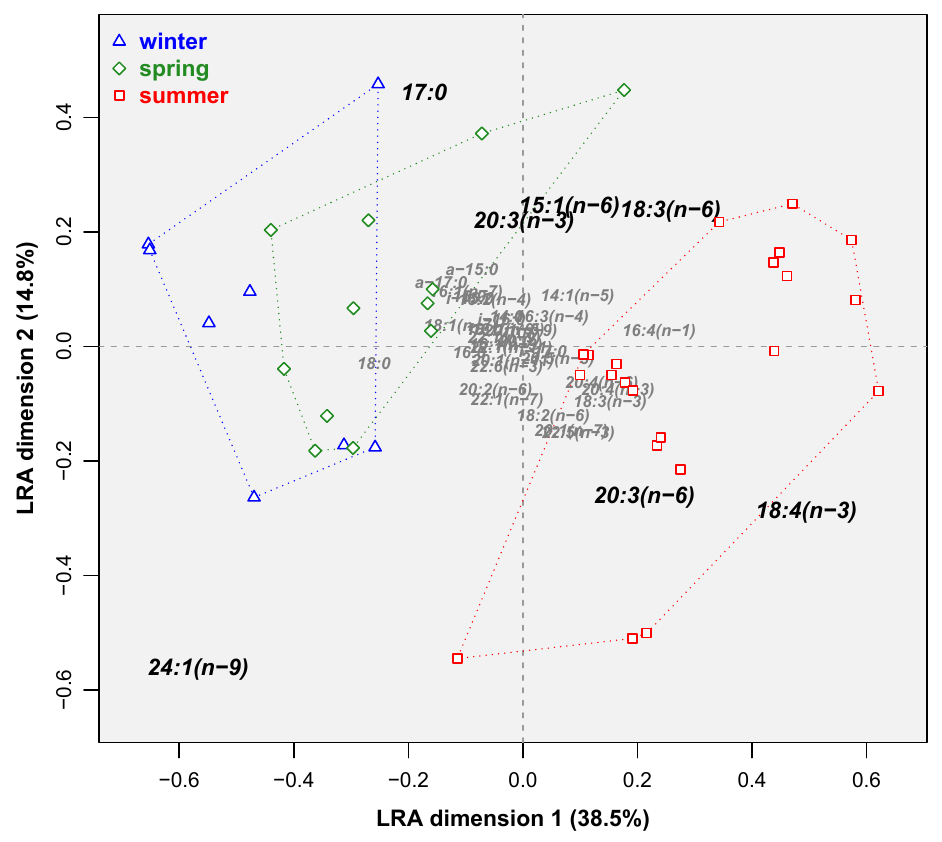}
\caption{LRA (i.e., PCA of the CLRs, equivalent ot PCA of all PLRs) of the original 40 FAs, showing the convex hulls of the three seasons' points, with overlap between winter and spring samples.Explained logratio variance is 53.3\% in this two-dimensional solution. Seven high-contributing FAs are shown in larger font, the remaining 33 FAs in smaller font.}
\label{FinalLRA}
\end{figure}
\end{center}

Finally, using the seven SLRs in Fig.~\ref{FinalResults}, a PCA  in  Fig.~\ref{FinalPCA} shows the two-dimensional solution, with the three seasonal groups well separated.
The seasons winter and spring are now clearly distinguished (see the overlap of winter and spring in Fig.~\ref{BarTernaryLRA}B and C, for example).
Winter is relatively higher on the PUFA and SFA amalgamation ratios pointing to lower left, and the spring samples higher on the MUFA amalgamation ratio pointing to the right. 
The ratios in the direction of winter can be explained by the start of the \emph{Calanus} using up reserves, which brings up more membrane bond FAs. 
In the spring direction the short-chain MUFAs are built up for storage during intensive feeding on algae under almost 24-hour daylight conditions.  

The seven SLRs, listed in Table \ref{SLRs}, explain 63.1\% of the total logratio variance. 
By contrast, the logratio analysis (LRA) in Fig.~\ref{FinalLRA} of all 40 FAs (i.e., the PCA of the 40 CLRs, equivalent to the PCA of all 780 PLRs, both of dimensionality 39), explains less variance, 53.3\%, on its first two dimensions.
This type of biplot \citep{AitchisonGreenacre:02} is often reported in the literature for a compositional data set, but in this case it does not show the clear separation of the seasons found in Fig.~\ref{FinalPCA}.
Moreover, there are 40 FA labels in Fig.~\ref{FinalLRA}, with many overlapping,  so the plot has been enhanced with FAs contributing more than average shown in larger font to improve legibility, whereas in Fig.~\ref{FinalPCA} there are only seven SLRs that were analysed.
This parsimonious set of SLRs, created using domain knowledge to define the amalgamations, has clearly been successful in all respects, with an excellent explanation of the seasonal differences.

%Finally, following \cite{Krzanowski:87}, \cite{Greenacre:18} has shown how Procrustes analysis \citep{Gower:04} can be used to measure similarity between two logratio geometries.
%The Procrustes correlation 

%There is an additional way of assessing the validity of the amalgamations and their logratios in the form of SLRs, namely to see how well they separate the seasons.
%There are two ways to do this: either use the seasons as a categorical variable explaining the SLRs, which can be done by applying RDA, or use the SLRs as predictors of the season categories in a trinomial logistic regression.
%The results are summarized in... {\color{red} [pending work for Michael]}.

%\setlength{\tabcolsep}{2.5pt}.
\begin{table}[t]
\vspace{-0.1cm}
\caption{Summary of the definitions of the amalgamations and the seven amalgamation (summated) logratios (SLRs) proposed as new variables for the analysis of fatty acids of marine organisms.}
\begin{tabular}{lll}
 {\sf\em Amalgamations}& {\sf\em Amalgamation logratios} & {\sf\em Abbreviations in Fig.~\ref{FinalPCA}}\\[2pt]
\hline\\[-6pt]
SFA, MUFA \& PUFA\ & log(PUFA/SFA) & \textsf{\textit{\smaller PUFA/SFA}}\\[1pt]
\  & log(MUFA/SFA) & \textsf{\textit{\smaller MUFA/SFA}}\\[2pt]
(PUFA) n3, n6 \& nX  & log(n6/n3) & \textsf{\textit{\smaller PUFAn6/n3}}\\[1pt]
\  & log(n6/nX) & \textsf{\textit{\smaller PUFAn6/nX}}\\[2pt]
(MUFA) short \& long chain & log(short/long) & \textsf{\textit{\smaller MUFAshort/long}}\\[3pt]
(SFA) 14:0\,+\,18:0 \& 16:0\,+\,20:0 & log($\frac{14:0+18:0}{16:0\,+\,20:0}$) & \textsf{\textit{\smaller SFA14+18/16+20}}\\[4pt]
(SFA) i-FAs\,+\,a-FAs \& 15:0\,+\,17:0  & log($\frac{\textrm{ i-FAs\,+\,a-FAs}}{{\rm 15:0\,+\,17:0}}$) & \textsf{\textit{\smaller SFA\,i+a/15+17}}\\[4pt]
\hline
\end{tabular}
\label{SLRs}
\vspace{-0.2cm}
\end{table}

\section{Discussion}
Our objective  has been to arrive at a small number of pairwise logratios of amalgamations (i.e., SLRs), where the amalgamations have substantive meaning for the practitioner (a biochemist, in this case) as well as being successful in explaining structural properties of the data set (the seasonal differences, in this case).

In a related work, \cite{Graeve:20} performed PLR selection, also step-by-step, on the same data set, with a combination of statistical and domain-knowledge criteria \citep{Greenacre:19}.
A list of PLRs that explained the highest percentages of logratio variance was determined statistically for the first step of the process. 
The biochemist selected one of these that had a clearly defined interpretation.
This PLR was fixed and then a second list was prepared, which explained the next highest percentage of variance, for similar consideration by the biochemist.
This process continued until six PLRs were chosen and deemed satisfactory.
%In that case, 91\% explained variance was achieved with only six PLRs.
However, those six PLRs were probably quite specific to the data set.
By contrast, in the present approach, the definition of the amalgamations can be considered more general and applicable as a benchmark set of variables.
Using ratios of sums of parts is also more reliable than using ratios of single parts.
Future studies of fatty acid compositions in marine biochemistry can use the set of seven amalgamation logratios (SLRs) proposed here to see how well they represent the compositional data structure.
The SLRs are defined by the horizontal pairwise links between the amalgamations in Fig.~\ref{FinalResults} and listed in Table \ref{SLRs} for future reference.

This idea of a hierarchy of amalgamations and their corresponding logratios can be exported to other fields, 
such as geochemistry where amalgamations of parts also occur naturally.  For example, \cite{GreenacreGrunskyBaconShone:20} analyse 10 major oxides of the Aar massif in Switzerland, nine of which subdivide into three amalgamations: mafic, felsic and carbonate-apatite. 
\cite{Grunskyetal:24} study diamond-bearing geological formations and define three sets of amalgamations of 22-part cation compositions: crustal, mantle and kimberlite. 
There is also a classification of the parts into major, minor, rare and trace elements. 
There is a potential for subdivisions of these groups of elements into substantively meaningful subsets, as performed with the fatty acids in the present work.  

The overall conclusion is that the construction of a hierarchy of amalgamated parts, based on domain knowledge of an expert, and their transformation to logratios, is a useful alternative way of performing variable selection in compositional data analysis.
The success of this selection can be quantitatively assessed using concepts and measures particular to this statistical field.

\section*{Author declarations}
The authors declare no competing interests. They also declare that they have used no AI tools in the preparation of this manuscript.

\section*{Data availability}
The fatty acid data set in this paper can be retrieved from 
\texttt{\smaller https://github.com/michaelgreenacre/CODAinPractice}
%\vspace{0.5cm}

\newpage

\bibliography{Amalgamations}

\end{document}